# ICT AND COMPETITIVENESS OF THE MACEDONIAN ECONOMY

**Olivera Kostoska,** *MSc in Economics*

**Pece Mitrevski,** *Associate Professor*

**Ilija Hristoski,** *MSc in Computer Science and Engineering Faculty of Economics - Prilep, St. Clement Ohridski University – Bitola Republic of Macedonia*

*"…as the economists are willing to point out, the most important is what happens "at the margins"-but the marginal technologies-ICT in particular-have an immense central and accelerating role in impelling the development processes"*
*Klaus Schwab, WEF*

ABSTRACT: The ongoing process of liberalization, as well as the intensive technological breakthrough have a great impact either on the macroeconomic policy creators, or the dissimilarity of conventional approach to found a strategy of a certain company. Bearing in mind the very notation, core characteristics of the concept of competitiveness are needed to be explored, but also the necessity to establish a national competitiveness strategy. However, the main attention should be paid on the technological changes that affect the economic growth, but the life standard increase, as well. Nevertheless, developing and transitional economies are being down-sided with respect to the technological and innovative capability that measurably implies the necessity to redefine the conception of competitiveness. Taking into account the above clarifications, the paper will set a comparative analysis of the key factors that pressure the competitiveness of Balkans region countries, mainly by the composition of the Growth Competitiveness Index. In addition, the major accent will be put on the composite Technology index and its impact on the mainstream of the respective total index. Finally, the paper will examine the possible change of Growth Competitiveness Index ranking of the Macedonian economy, principally by the variation of the Technology index rank.

## 1. INTRODUCTION

Nowadays, improvement of the Information and Communication technology (ICT) sector has progressively decreased the costs imposed by managing information, thus facilitating individuals and organizations to undertake information-related tasks much more efficiently. There is widespread evidence that a great portion of the productivity growth in the world's leading economies is being compelled by the impact of their particular ICTs. Namely, investment in ICT and technical progress accounted for approximately 40% of European Union (EU) labor productivity growth within the period of 90s, whereas it accounted for about 60% in the USA [1]. The importance of ICT has been widely recognized by the European Heads of State and Government meeting held in Lisbon, in March 2000, wherein the objectives to make EU the most competitive knowledge-based economy were set down. Nonetheless, the so called "Lisbon strategy" was established to achieve sustainable economic growth of the EU aimed at obtaining more and better jobs, as well as greater social cohesion. The comprehensive series of interdependent and self-reinforcing reforms proposed by the very Strategy stand for the eight distinctive dimensions, such as [2]: creating an information society for all (which is the objective of the European Action Plan), developing an European area for innovation, Research and Development (R&D), liberalization (completing the single market, state aid and competition policy), building network industries (in telecommunications, utilities and transportation), creating efficient and integrated financial services, improving the enterprise environment (particularly in terms of regulatory framework), increasing social inclusion, enhancing sustainable development.

Despite the above mentioned economies, Republic of Macedonia is being still led by the traditional factors of production i.e. the country counts upon the cheap labor force and natural resources exploitation that stipulates a kind of price-based competitiveness instead of a structural one. To be exact, macroeconomic stability is merely the factor to be used in upgrading the competitiveness at a global level vis-à-vis ICT and technological readiness that should be tackled upon in the years ahead. In other words, the Growth Competitiveness Index suggests that Republic of Macedonia transits from price competitiveness economy to the one based on qualitative products and effective markets. This is to be associated with the time stretch needed to achieving the EU standards, as well as to make the country compatible with the EU regulative and the global competitiveness [3].

## 2. KEY ELEMENTS OF SUSTAINABLE COMPETITIVENESS

ICTs have rapidly abraded the space-time distance between the countries, making them more integrated, particularly observed throughout the cross-border trade, investment and financial flows. In principle, liberalization, as well as the open technological changes has a great impact either on the national or the international aspects while one creates the trade policy of a certain economy. Owing to this, the national strategic disposure is supposed to be shaped considering the so called *national competitive strategies*. In accordance with the World Economic Forum (WEF), national competitiveness is defined as a set of factors, policies and institutions that determine the productivity level of a particular country. The most common view of competitiveness is being perceived as an enhanced export performance of a certain economy, deliberated mostly by the upward market share within the world economy. However, it should be emphasized that numerous of other factors affect the national competitiveness, such as: macroeconomic stability, but more important, property rights (De Sotto, 2000), openness and transparency in the management of public recourses (Kaufmann, 2005), education and training, fiscal management etc. Finally, beyond the above factors, latest technologies, the ICT in particular, play the major role in enhancing the productivity growth i.e. ICT is being considered as a key driver of the national economy competitive growth.

As of 2001, the Growth Competitiveness Index (GCI) developed by Jeffrey Sachs and John McArthur is being utilized by WEF in order to assess the competitiveness of the nations. At the very commencement of setting up, the Index was observed as an elegant endeavor to comprise a huge number of factors that affect the productivity within many countries. However, lots of annotations about the particular Index have emerged recently, principally imposed by its major limitation, not to include any indicator that would comprise the efficiency of labor market. From the other side, the above mentioned Lisbon strategy launched to place the EU into the most competitive region by 2010 emphasizes the efficiency of the labor market as a main prerequisite of the productivity growth. Many other indicators also were not included within the stated Index, such as: public health, modernization of the country's infrastructure etc. In order to overcome those weaknesses, professor Sala-i-Martin has developed a brand new wide-ranging so called Global Competitiveness Index, firstly introduced in the WEF Global Competitiveness Report 2004-2005 (Sala-i-Martin and Artadi, 2004) [4]. Nonetheless, this paper analyses will be carried out by utilization of the composite Growth Competitiveness Index due to the more extensive time series obtained by the heretofore WEF Reports considering the Balkans region countries.

The Growth Competitiveness Index is composed of three fundamental components [5]: technology index, public institutions index and macroeconomic index. These indexes are computed on basis of a sample of countries considered to be core and non-core innovators, as well as two data categories (hard and survey ones). The weightings of the specific indexes within the total one for the core innovators are given by the following equation:

$$GCI = \frac{1}{2} \cdot TI + \frac{1}{4} \cdot PII + \frac{1}{4} \cdot MEI \qquad (1)$$

where TI stands for Technology index, PII denotes Public institutions index, while MEI symbolizes the Macroeconomic environment index, while indexes for the non-core innovators are equally included i.e. GCI represents a simple average of the very three components:

$$GCI = \frac{1}{3} \cdot TI + \frac{1}{3} \cdot PII + \frac{1}{3} \cdot MEI \qquad (2)$$

Yet, each component index is being consisted of several sub-indexes. Thus, the Technology index (TI) for core innovators is computed as follows:

$$TI = \frac{1}{2} \cdot IS + \frac{1}{2} \cdot ICTS \qquad (3)$$

where IS stands for Innovation sub-index, whilst ICTS denotes Information and Communication Technology sub-index. The sub-indexes' ponders for the non-core innovators are given by the equation below:

$$TI = \frac{1}{8} \cdot IS + \frac{3}{8} \cdot TTS + \frac{1}{2} \cdot ICTS \qquad (4)$$

where TTS stands for Technology transfer sub-index.

The components of Public institutions index (PII) are added in an even way i.e.

$$PII = \frac{1}{2} \cdot CLS + \frac{1}{2} \cdot CS \qquad (5)$$

where CLS symbolizes the Contracts and Law sub-index, while CS denotes the Corruption sub-index.

Finally, the Macroeconomic environment index (MEI) is composed by three sub-indexes, disposed as follows:

$$MEI = \frac{1}{2} \cdot MSS + \frac{1}{4} \cdot CCR + \frac{1}{4} \cdot GW \qquad (6)$$

where MSS denotes the Macroeconomic stability sub-index, CCR stands for Country credit rating and GW symbolizes the Government waste.

Considering the fact that focal point of this paper is the impact of ICT on competitiveness of the national economy (Macedonian economy, within the case in point), we shall make a comparative preview of the overall GC index at first, and obtain more detailed analysis of the Technology index and ICT sub-index (ICTS) as a constituent of the previous one. In addition, ICT sub-index is being computed, as follows:

$$ICTS = \frac{1}{3} \cdot ICTsd + \frac{2}{3} \cdot ICThd \qquad (7)$$

where ICTsd symbolizes ICT survey data, whilst ICThd stands for ICT hard data. In addition, it has to be noted that data from ICT survey is being obtained by the following questions: Internet access in schools, Quality of competition in the ISP[55] sector, Government prioritization of ICT, Government success in ICT promotion and Laws relating to ICT, while ICT hard data refers to the number of Cellular telephones, Internet users, Internet hosts, Telephone lines and Personal computers.

## 3. COMPARATIVE ANALYSIS OF GCI, TECHNOLOGY INDEX AND ICT SUB-INDEX WITHIN THE BALKANS REGION

By virtue of their distinctive histories, geography, and social conditions, countries are at widely varying levels of income, technological sophistication, capacity to innovate, and overall capacity to achieve sustained economic growth. Bearing in mind the very notation, the Growth Competitiveness Index (GCI) appraises the capability of a certain country to reach a sustainable economic growth on a medium term. However, it should be noted that the growth prospects of a particular economy depends not only upon the GCI scores, but also the level of GDP per capita. In other words, a deprived country with a low GCI score might still have good growth prospects comparative to a richer one. Furthermore, GCI ranking should not be confounded with the GCI score i.e. the second one could be found within the scale of 1-7. Namely, the statistical evidence illustrates that an increase of 1 point in GCI score for two economies with the same GDP per capita income, is being related on average to an insignificant rice of the growth rate of more than 3 percentage points annually. Moreover, geographical factors are not directly taken into account within the Index composition, since those affect industrial structure and other variables already included in the GCI. Finally, the Index does not account for the internal variations of the particular country, thus it should be not confused with the worst growth prospects [6].

In this context, we will assess the GCI position of the Balkans countries using data[56] from the recent years. To be exact, our focal point is being motivated due to several reasons, such as: Republic of Macedonia is an inseparable part of this region, the Balkans countries are being found at different stages of economic development and they also have unlike status[57] with respect to the EU integration processes.

---

[55] Internet Service Provider

[56] Data used in this paper is obtained from the WEF Global Competitiveness Reports 2001-2006.

[57] Currently (2007), Greece, Romania, Bulgaria and Slovenia are already EU members, Macedonia, Croatia and Turkey are considered as EU candidates, while Serbia and Montenegro, Albania and Bosnia and Herzegovina stand for potential EU candidate countries. Even though Serbia and Montenegro subsist like independent countries at the moment, they will be represented as a group within the paper due to the WEF data available.

Figure 1. GCI scores about Balkans countries (2001-2006)

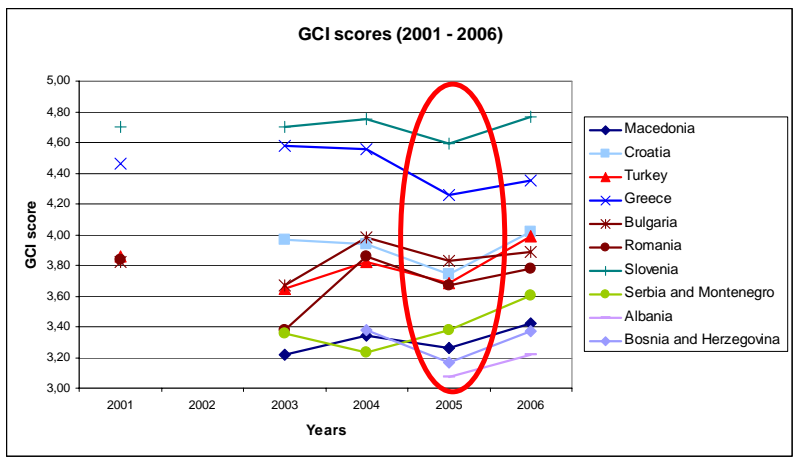

Source: WEF Global Competitiveness Reports 2001-2006

The records of GCI scores for the selected countries (Figure 1), during the period 2001-2006, illustrate that all the Balkans countries (except EU members i.e. Slovenia and Greece) are being sited approximately at the same level that is correlated with their individual GDP/capita rates. However, the evidence presented indicates that the GC index significantly differs among the pointed countries in 2006, whereupon Slovenia, Greece and Croatia in addition, capture the highest GCI scores within the region (4.77, 4.35 and 4.02, respectively).

Figure 2. GCI rankings of the Balkans countries (2005-2006)

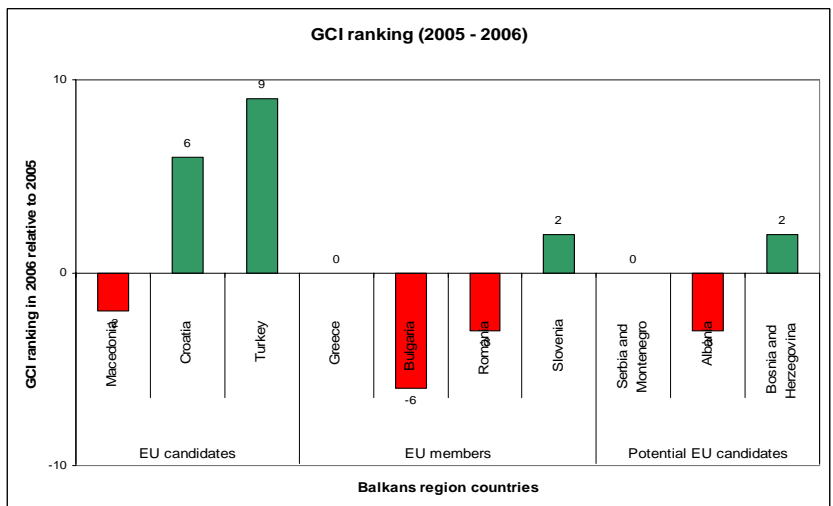

Source: WEF Global Competitiveness Reports 2005-2006

Nonetheless, some noteworthy evidence provided from the Figure 1 shows that considerable fall of GCI has emerged amid all the countries in 2005, mainly imposed by the slowdown in industrial production and foreign trade, as well as the adverse impact of higher oil prices [7]. However, according to GCI Republic of Macedonia is being found at the bottom of the region during the entire period (for instance, Bosnia and Herzegovina and Albania are the only ones scored below the Macedonian economy in 2006). On the other hand, the highest GCI ranking rise in 2006 relative to 2005 (Figure 2) has Turkey (+9) and Croatia (+6), but the uppermost fall within the same period has Bulgaria (-6). Republic of Macedonia is the single one among EU candidates that performs relative fall of two positions.

Table 1. Chi-Square hypothesis testing
(PHStat for MS Excel)

| Results | |
|---|---|
| Critical Value | 16,91898 |
| Chi-Square Test Statistic | 1,459644 |
| *p*-Value | 0,997435 |
| Do not reject the null hypothesis | |

As one of the EU members taken into account in the paper analysis, Slovenia is ranked two places higher in 2006 relative to 2005, while Greece retains the very same position.

Although different variations are presented in GCI rankings for a particular country, there is no statistically significant dissimilarity in the overall ranking of the Balkans region as a whole during the preceding two years. Namely, the p-value (0,997435) of the Chi-Square test statistics computed at the level of significance of alpha=5% indicates that we should not reject the null hypothesis (p>alpha). Thereby the above mentioned claim is being confirmed (Table 1).

However, it is worthy to mention that the average GCI score of the Macedonian economy for the duration of 2003-2006 is being increased (Figure 3). In other words, the competitive position of the country is being amended due to a number of parameters, but decreased as a consequence of some others. Namely, positive impact on augmented average GCI score has the macroeconomic stability of the Macedonian economy, mainly imposed by the appropriate mixture of the monetary, fiscal and income policy effectuated inside the country. As a matter of fact, the only one steady improvement of Macedonian competitiveness could be discovered within the frame of macroeconomic environment.

Nevertheless, the above state is determined by the phase wherein the country is being found i.e. Republic of Macedonia transits from the competitiveness based on fundamental factors of production (the competitive advantages are gained from the contemptible factors endowment) towards the second phase (efficiency based competitiveness). On the other hand, radical fall of the Technology index is being inflicted by not taking important measures aimed at improving the country's technological readiness, principally when the other economies have advancement in this domain. Thus, time series linear regression of the Technology index indicates that each year the average score of particular Index is being decreased for 0,242 units, despite the GCI regression where the score is being amplified in about 0,052 units per annum. In addition, the Pearson correlation coefficient between the two variables takes a negative value (-0,39119).

The necessity to analyze the ICT sub-index emerges from its importance for the prevalent know-ledge based economy, primarily pressuring the quality and price of a certain country estimated by the level of its exports. In other words, ICT has a great impact on the growth and total factor productivity (TFP) of the particular economy no matter if it is imported or internally developed.

Figure 3. GCI and Technology index

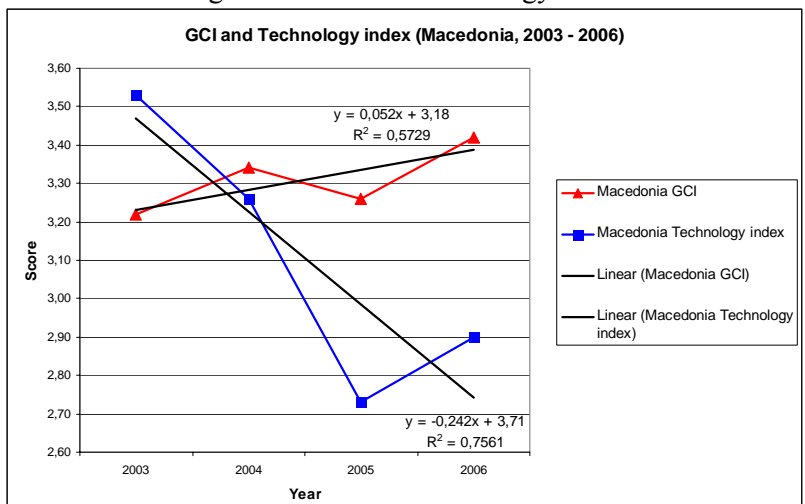

Source: WEF Global Competitiveness Reports 2003-2006

The utilization, as well as the ICT improvement is considered to be critically important, not because of the enormous necessity to establish fast and effective system of communications, but also to obtain an efficient infrastructure intended for commercial transactions.

Figure 4. ICT sub-index for the Balkans region

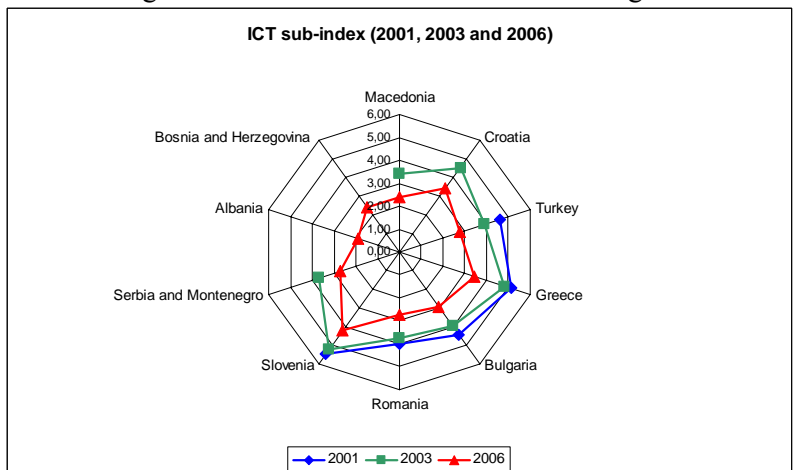

Source: WEF Global Competitiveness Reports 2001-2006

Notwithstanding, Republic of Macedonia has very low scores of the ICT sub-index i.e. the country has a technological lag in all the components of the particular sub-index, except for the usage of mobile phones and personal computers. In addition, Macedonian economy does not have either an appropriate legislation regarding to the ICT domain (e-commerce law) or proper implementation of the adopted one (digital signature). The evidence presented above (Figure 4) shows that Republic of Macedonia falls behind the countries of the region (Albania is an exception), but all of them stand behind the mayor number of the EU countries.

In other words, there is a notable fall of the ICT sub-index despite the GC index within the entire Balkans region during the period 2001-2006. This fact suggests that the ICT environment in these countries is not convenient enough. Moreover, the foremost aberrations could be found in the Internet access, the exploitation of personal computers and mobile phones, as well as the appliance of the new IC technologies at a firm level.

## 4. CONCLUSIONS

Technological dissimilarity between the countries is considered to be the main reason for the differentiation in the levels of productivity. During the last few decades, new technologies have become the major prerequisite in competitiveness augmentation, mainly imposed by the improvement and more frequent utilization of the ICT. As a matter of fact, technological readiness of a certain economy is being usually estimated by means of ICT availability, as well as the stridency of its usage. In addition, technologically-intensive foreign direct investments do not improve the productivity only, but impose some additional side-effects (acquirement new skills for the employees). Unfortunately, Republic of Macedonia notes unsatisfactory results in each domain of technological readiness. As the evidence shows, the worst records are to be found among the factors which have an important impact on the growth of the country (FDI and technology transfer, ICT and legislation, in particular). Nevertheless, the Balkans region as a whole could not be considered as one with high level of technological readiness, except for some countries (Croatia) which also fell notably behind the EU members. Bearing in mind the impact of ICT on the competitiveness of a particular economy, Republic of Macedonia should prepare and effectuate an appropriate policy for technological development aimed at increasing the R&D activities as the main instigator of the new technologies and innovations. In other words, improvements in the ICT sector are not considered necessary to establish fast and effective system of communications only, but also to obtain an efficient infrastructure intended for commercial transactions.

## 5. BIBLIOGRAPHY

[1] Reding, V., "i2010: The European Commission's new programme to boost competitiveness in ICT sector", *Microsoft's Government Leaders Forum,* Prague, 2005
[2] Chevallerau, F., "The impact of E-Government on competitiveness, growth and jobs", *Background Research Paper*, *IDABC eGovernment Observatory*, 2005
[3] Center for Economic Analyses, *" Institutions as determinants of economic growth"*, *National Council for Entrepreneurship and Competitiveness*, Skopje, 2006
[4] Lopez-Claros, A. et al., "The Global Competitiveness Index: Identifying the Key Elements of Sustainable Growth", *World Economic Forum 2006-2007*, pp. 3-51, 2007
[5] McArthur, J. and Sachs, J., "Growth Competitiveness Index: Measuring Technological Advancement and the Stages of Development", *World Economic Forum 2001-2002,* pp. 28-51, 2002
[6] World Economic Forum, "*Global Competitiveness Report 2001-2002"*, Geneva, pp. 28-50, 2002
[7] World Economic Forum, "*Global Competitiveness Report 2005-2006"*, Geneva, 2006